\documentclass{emulateapj}

\usepackage{graphicx}
\usepackage{graphics}

\def\gsim{\;\lower4pt\hbox{${\buildrel\displaystyle >\over\sim}$}\,}
\def\lsim{\;\lower4pt\hbox{${\buildrel\displaystyle <\over\sim}$}\,}

\def\FLASH{{\sc flash}}
\def\PARAMESH{{\sc paramesh}}

\newcommand\referee[1]{{ #1}}

\slugcomment{}

\shorttitle{Ejecta clumping and back-reaction of accelerated cosmic
   rays in SNRs}
\shortauthors{S. Orlando et al.} 

\begin{document}

\newcommand\rs[1]{_\mathrm{#1}}
\newcommand\op[1]{{\bf #1}}

\title{Role of ejecta clumping and back-reaction of accelerated cosmic
          rays in the evolution of Type Ia supernova remnants}

\author{S. Orlando\altaffilmark{1}, F. Bocchino\altaffilmark{1},
        M. Miceli\altaffilmark{2,1}, O. Petruk\altaffilmark{3,4},
        M.L. Pumo\altaffilmark{5,6}}
\email{orlando@astropa.inaf.it}

\altaffiltext{1}{INAF - Osservatorio Astronomico di Palermo ``G.S.
              Vaiana'', Piazza del Parlamento 1, 90134 Palermo, Italy}
\altaffiltext{2}{Dip. di Fisica, Univ. di Palermo, Piazza del Parlamento 1, 90134 Palermo,
              Italy}
\altaffiltext{3}{Institute for Applied Problems in Mechanics and
              Mathematics, Naukova St. 3-b Lviv 79060, Ukraine}
\altaffiltext{4}{Astronomical Observatory, National University, Kyryla and
              Methodia St. 8 Lviv 79008, Ukraine}
\altaffiltext{5}{INAF -- Osservatorio Astronomico di Padova, Vicolo
              dell'Osservatorio 5, 35122 Padova, Italy}
\altaffiltext{6}{INAF -- Osservatorio Astrofisico di Catania, Via S. Sofia
              78, 95123 Catania, Italy}

\begin{abstract}
We investigate the role played by initial clumping of ejecta and by
efficient acceleration of cosmic rays (CRs) in determining the density
structure of the post-shock region of a Type Ia supernova remnant (SNR)
through detailed 3D MHD modeling. Our model describes the expansion of
a SNR through a magnetized interstellar medium (ISM), including
the initial clumping of ejecta and the effects on shock dynamics due to
back-reaction of accelerated CRs. The model predictions are compared
to the observations of SN 1006. We found that the back-reaction of
accelerated CRs alone cannot reproduce the observed separation between
the forward shock (FS) and the contact discontinuity (CD) unless the
energy losses through CR acceleration and escape are very large and
independent of the obliquity angle. On the contrary, the clumping of
ejecta can naturally reproduce the observed small separation and the
occurrence of protrusions observed in SN 1006, even without the need of
accelerated CRs. We conclude that FS-CD separation is a probe of the
ejecta structure at the time of explosion rather than a probe of the
efficiency of CR acceleration in young SNRs.
\end{abstract}

\keywords{cosmic rays --- 
          magnetohydrodynamics (MHD) --- 
          instabilities --- 
          shock waves --- 
          ISM: supernova remnants --- 
          supernovae: individual: SN 1006}

\section{Introduction}
\label{sec1}

Nowadays it is widely accepted that supernova remnants (SNRs) are the site
where cosmic ray (CR) diffusive shock acceleration occurs. Observations in
various bands support this picture through the detection of non-thermal
emission that is compatible with being synchrotron or Inverse Compton
radiation from CR electrons. Unfortunately, the direct evidence of
CR ions in SNRs is difficult to find because they do not radiate
efficiently. On the other hand, different indirect signatures of the
presence of CR ions are largely discussed in the literature. The
most popular is probably the separation between the forward shock
and the contact discontinuity that has been measured in young SNRs
(e.g. SN\,1006, \citealt{2009A&A...501..239M}, and Tycho's SNR,
\citealt{2005ApJ...634..376W, 2007ApJ...665..315C}). In fact, current
theories predict that a significant fraction of the energy of supernova
remnant shocks is channeled into CRs, determining modifications of the
shock dynamics that depend on the efficiency of acceleration and injection
processes of high energy particles. In particular, this energy losses
would lead to a greater shock compression ratio and, as a consequence,
to a thinner shell of shocked interstellar medium (ISM).

An example of SNR in which the observed features have been interpreted
as a consequence of the energy losses to CRs at the forward shock
is SN\,1006. In this remnant, the observations have shown that the
azimuthal profile of the ratio of the forward shock radius to the contact
discontinuity radius $R\rs{fs}/R\rs{cd}$ is fairly uniform (although
very noisy) and much lower than predicted for a non-modified shock
(\citealt{2009A&A...501..239M}). Recently \citet{2011ApJ...735L..21R}
have found and analyzed clumps of ejecta close to or protruding beyond the
main blast wave of SN\,1006 that have been interpreted in the context of
an upstream medium modified by the saturated nonresonant Bell instability
which enhances the growth of Rayleigh-Taylor (RT) instabilities at the
contact discontinuity.

However, some pieces of evidence are now accumulating that are
difficult to explain in terms of acceleration of CR particles. Some
authors (e.g. \citealt{2001ApJ...560..244B, 2011MNRAS.415...83W} and
references therein) noted that extreme energy losses to accelerate the
CRs are needed to allow a significant fraction of the ejecta to approach
or even overtake the forward shock, thus explaining the thin shell of
shocked ISM. \cite{2011MNRAS.415...83W} analyzed the evolution of RT
instabilities in Type Ia SNRs undergoing CR particle acceleration and
found that, even with very efficient acceleration of CRs (i.e. assuming
an effective adiabatic index $\gamma\rs{eff}\approx 1.1$), significantly
enhanced mixing and perturbation of the remnant outline are not
expected. A similar conclusion was reached by \cite{2010A&A...515A.104F}
who found that the development of RT instabilities in SNRs is not
drastically affected by CR particle acceleration. In addition, these
studies suggest that the high occurrence of protrusions in young SNRs
is not the consequence of RT instabilities enhanced by accelerated CRs
(see also \citealt{2001ApJ...549.1119W}). Another evidence difficult
to explain in terms of acceleration of CR particles is the ratio
$R\rs{fs}/R\rs{cd}$ measured in SN\,1006 that is lower than predicted
by non-modified shock models even in regions dominated by thermal
emission where the CR acceleration efficiency is supposed to be low
(e.g. \citealt{2009A&A...501..239M}). All these studies cast some
doubts on whether the back-reaction of accelerated CRs is the main
responsible of the enhanced intershock instabilities observed in young
SNRs (e.g. SN\,1006 and Tycho's SNR).

On the other hand, spectropolarimetric studies of SNe Ia show the
presence of asymmetries with different magnitude and orientation for
different elements in the ejecta and the detection of strong line
polarization (e.g. \citealt{2003ApJ...591.1110W, 2004ApJ...604L..53W,
2005ApJ...632..450L, 2006ApJ...653..490W, 2008AJ....136.2227C,
2010ApJ...720.1500H}). All these features have been interpreted as being
due \referee{to} clumpy structures in the outer layers of the ejecta
(see \citealt{2010ApJ...720.1500H} and references therein) and some
authors suggested that ejecta clumps of intermediate-mass elements can
be forged in the explosion of SN Ia (e.g. \citealt{2003ApJ...591.1110W,
2005ApJ...632..450L}) or may be due to the interaction of the ejecta
with a dense, clumpy, and disk-like circumstellar environment (e.g.
\citealt{2004ApJ...604L..53W}). Recently \cite{2010Natur.466...82M}
have shown that asymmetries in the explosion can be a generic feature in
SNe Ia (see also \citealt{2010ApJ...712..624M}), and these asymmetries,
in turn, may lead to a clumpy structure of the ejecta. In the light of
these considerations, it is therefore important to investigate the role
of ejecta clumping on the evolution and morphology of Type Ia SNRs. In
particular we wonder whether the thermal and density structure of the
post-shock region of a young SNR originates mainly from the clumpy
structure of the ejecta rather than as a consequence of back-reaction
of accelerated CRs.  The density inhomogeneities in the ejecta can
enhance the growth of RT instabilities, causing the ejecta material to
move closer to the main blast. The question is: can the ejecta clumping
enhance the growth of RT instabilities up to a level that allows clumps
of ejecta to reach and possibly overtake the forward shock?

Here we investigate this issue by developing a three-dimensional (3D)
MHD model describing the expansion of a SNR through a magnetized medium,
including, for the first time, the (non-uniform) ambient magnetic field,
the initial ejecta clumping, and the effects on shock dynamics due to
back-reaction of accelerated CRs. The paper is organized as follows:
in Sect.~\ref{sec2} we describe the MHD model and the numerical setup;
in Sect. \ref{sec3} we describe the results and, finally, we draw our
conclusions in Sect. \ref{sec5}.

\section{MHD model and numerical setup}
\label{sec2}

The evolution of a SNR can be characterized by distinct stages
depending on the physical process dominating its dynamics
(e.g. \citealt{1977ARA&A..15..175C}). This paper focuses on young
SNRs, i.e. remnants that have evolved from the ejecta-dominated stage
through the Sedov-Taylor stage. Pioneering comprehensive studies of
the dynamics of these remnants, preceding the onset of dynamically
significant radiative losses and/or pressure confinement by the ambient
medium, are given in the literature (e.g. \citealt{1974ApJ...190..305M,
1994ApJ...435..805F, 1999ApJS..120..299T}) and are mostly based on
analytic and numerical 1D hydrodynamic models. Subsequently several
2D and 3D hydrodynamic and MHD models describing the evolution of the
remnant through the ISM have been developed.

Here we adopted the 3D MHD model discussed by
\citet{2007A&A...470..927O, 2011A&A...526A.129O}, extended to describe
the initial ejecta clumping and to include the effect of larger
compressibility of plasma around the shock due to the back reaction
of accelerated CRs. The shock propagation is modeled by numerically
solving the time-dependent ideal MHD equations of mass, momentum,
and energy conservation in a 3D cartesian coordinate system $(x,y,z)$
(see \citealt{2007A&A...470..927O} for details). In order to trace the
motion of the ejecta material and study its dynamics, we considered a
passive tracer associated with the ejecta. The continuity equation of
the tracer is solved in addition to our set of MHD equations; the ejecta
material is initialized with $C\rs{ej} = 1$, while $C\rs{ej} = 0$ in the
ISM. The calculations were performed using \FLASH\ (\citealt{for00}),
an advanced multi-dimensional MHD code for astrophysical plasmas,
including the adaptive mesh refinement through the \PARAMESH\ library
(\citealt{mom00}), and extended with additional computational modules
to handle the back-reaction of accelerated CRs.

The effects of shock modification are included in the MHD model by
following the approach of \citet{2010A&A...509L..10F} and extending their
method to MHD models. In particular, our model includes an effective
adiabatic index $\gamma\rs{eff}$ which depends on the injection rate
$\eta$ of particles (i.e. the fraction of ISM particles entering
the shock front). The adiabatic index on the shock is varied due
to particle acceleration as in \cite{2004A&A...413..189E} (see also
\citealt{2010A&A...509L..10F}). At each time-step of integration, the
adiabatic index is calculated at the shock front and then is advected
within the remnant, remaining constant in each fluid element. As discussed
by \citet{2010A&A...509L..10F}, the latter assumption implies that each
fluid element remembers the effect of shock modification induced by
particle acceleration at the time it was shocked.

For the purposes of the present paper, we assume that the maximum
injection rate $\eta$ is large enough (e.g. $\eta \approx 10^{-3}$,
namely when shock modifications are strong and immediate) so that the
effective adiabatic index at the initial conditions of our simulations
has already reached its minimum value and slightly depends on time
(\citealt{2010A&A...509L..10F}). We assume therefore the effective
adiabatic index not depending on time and consider its minimum value
$\gamma\rs{min}$ as a free parameter. On the other hand, the injection
rate is expected to depend on the shock obliquity (i.e. the angle between
the unperturbed external magnetic field and the normal to the shock;
e.g. \citealt{2003A&A...409..563V}). We allow therefore that the effects
of shock modification on the fluid dynamics (and, therefore, the effective
adiabatic index) vary in space as a function of the obliquity angle. We
assume no magnetic field amplification due to CRs, and no back-reaction
of accelerated CRs at the reverse shock, although the suggestion that
CR particles can be efficiently accelerated also at the reverse shock
is largely debated in the literature (e.g. \citealt{2005A&A...429..569E}).

The index $\gamma\rs{eff}$ is calculated at the shock front by using a
parametrized function depending on the obliquity angle $\Theta\rs{o}$
and characterized by a parameter representing the minimum value of the
adiabatic index $\gamma\rs{min}$ that is possible to reach during the
simulation:

\begin{equation}
\gamma\rs{eff} = \gamma-\left(\gamma-\gamma\rs{min}\right)
\times f\rs{\varsigma}(\Theta\rs{o})
\label{eq1}
\end{equation}

\noindent
where $\gamma = 5/3$ is the adiabatic index and $f\rs{\varsigma}
(\Theta\rs{o})$ is a function defined in the range $[0,1]$ depending on
the obliquity angle $\Theta\rs{o}$ and describing the variations of
$\gamma\rs{eff}$ over the surface of the remnant shock. In analogy
with the description given by \cite{1990ApJ...357..591F} for the
quasi-parallel, quasi-perpendicular, and isotropic injection models
(see also \citealt{2007A&A...470..927O, 2011A&A...526A.129O}), we
model the variations of $\gamma\rs{eff}$ over the shock surface
through the functions $f\rs{\varsigma}(\Theta\rs{o}) = \cos^2
\Theta\rs{s}$ (i.e. $\gamma\rs{eff}$ is minimum at parallel shocks),
$f\rs{\varsigma}(\Theta\rs{o}) = \sin^2 \Theta\rs{s}$ ($\gamma\rs{eff}$
is minimum at perpendicular shocks), and $f\rs{\varsigma}(\Theta\rs{o})
= 1$ ($\gamma\rs{eff}$ is uniform at the shock front and equal to its
minimum value), where $\Theta\rs{s}$ is the angle between the shock
normal and the post-shock magnetic field and is related to $\Theta\rs{o}$
by the expression $\cos \Theta\rs{s} = \sigma^{-1} \cos \Theta\rs{o}$
and $\sigma$ is the shock compression ratio. The first case follows
the quasi-parallel injection scenario, leading to a 3D polar-caps
structure of the remnant, whereas the second and the third cases follow
the quasi-perpendicular and isotropic injection models, respectively,
producing a 3D equatorial-belt structure of the remnant. Note that the
third case (isotropic) is intended to be the extreme case in which the
shock modification is the largest everywhere at the forward shock with
no obliquity dependence.

\begin{figure}[!t]
  \centering
  \includegraphics[width=8.5cm]{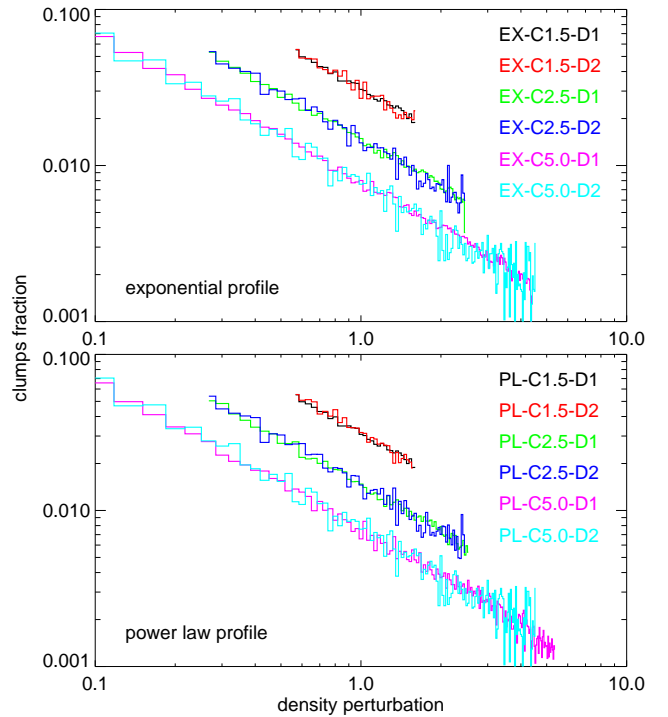}
  \caption{Probability distribution functions of the random perturbation
           of mass density of the clumps for the two ejecta density
           profiles considered in this paper: the exponential profile
           (upper panel) and the \referee{power-law} profile with index $n=7$ (lower
           panel). The density perturbation of each clump is calculated as
           the ratio of the density of the resulting \referee{clump to
           the local average density in the region occupied by the clump
           if the perturbation was not present.}}
  \label{fig1}
\end{figure}

As for the density structure of the ejecta, we investigated: the
exponential profile that has been shown to be the most representative of
explosion models for thermonuclear SNe (\citealt{1998ApJ...497..807D}),
and the power-law profile with index $n=7$ that has been used
to represent deflagration models (\citealt{1983ApJ...272..765C,
1984ApJ...286..644N}). We also assume that the initial ejecta has a
clumpy structure. The clumps have been modelled as per-cell random
density perturbations\footnote{The density perturbation of each
clump is calculated as the ratio of the mass density of the resulting
\referee{clump to the local average density in the region occupied
by the clump if the perturbation was not present.}} derived from a
\referee{power-law} probability distribution\footnote{\referee{Since no
observational clues are available on the distribution of density perturbations
of the clumps, it is reasonable to assume that most of the clumps are
characterized by small density perturbations and few of them by large
perturbations. To this end, for the sake of simplicity, we assume a
power-law probability distribution with index $n = -1$.}} (index $n =
-1$) that is characterized by a parameter $\nu\rs{max}$ representing the
maximum density perturbation allowed in the simulation. Figure~\ref{fig1}
shows the \referee{power-law} probability distributions of the
perturbations used in this paper for the two ejecta density profiles
considered. We explored maximum density perturbations ranging between
1.5 and 5; we explored density clumps of ejecta with size either $1$\% or
$2$\% of the initial diameter of the remnant $D\rs{snr0}$. As discussed
in Sect.~\ref{sec:clump}, initial clump size in the range explored here
leads, after 1000~yr of evolution, to density features with characteristic
size comparable to those observed in SN\,1006.

\begin{figure}[!t]
  \centering
  \includegraphics[width=8.5cm]{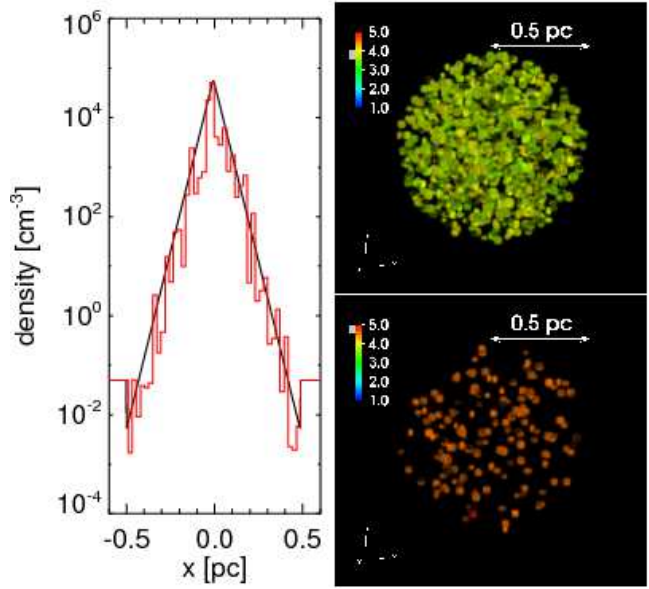}
  \caption{Left panel: initial spatial distribution of plasma density
           along the $x$-axis for a model either with (red line; run
           EX-C5.0-D2 in Table~\ref{tab1}) or without (black line; run
           REF-EXP) the ejecta clumping. \referee{In both models the
           total mass of ejecta (integrated over the whole volume) is
           $1.4~M_{\rm sun}$ (see text).} Right panels: initial spatial
           distributions of ejecta clumps with density perturbation
           in the range either $[3.5-4]$ (upper panel) or $[4.5-5]$
           (lower panel) in run EX-C5.0-D2.}
  \label{fig2}
\end{figure}

It is interesting to note that the range of clump size
investigated in this paper is also in agreement with that derived by
\cite{2010ApJ...720.1500H} for SNe Ia. In particular these authors
compared the results of their semi-analytic code for modeling polarized
line radiative transfer within 3D inhomogeneous rapidly expanding
atmospheres with spectropolarimetric observations; they found that the
model reproduces the observed range of values of peak line polarization
if the clumps have radius in the range\footnote{Sizes are given in
velocities because of the linear dependence of velocity on distance in
the homologous flow of SN ejecta (this structure is sometimes referred
to as a pseudo-Hubble flow; see \citealt{2010ApJ...720.1500H}).}
$1000-6000$~km~s$^{-1}$. At the time of our initial condition ($\approx
10$~yr since the SN explosion), the effective range of clump size derived
by \cite{2010ApJ...720.1500H} corresponds to $0.016-0.13$~pc to be
compared with the size of the clumps modelled here ranging between 0.01
and 0.02~pc. As an example, Fig.~\ref{fig2} shows the initial spatial
distribution of ejecta clumps for a model with the highest density
perturbation and largest clump size. A summary of all the simulations
discussed in this paper is given in Table~\ref{tab1}.

\begin{table*}
\footnotesize
\caption{Adopted parameters and initial conditions for the MHD
models of the SNR
\label{tab1}}
\begin{center}
\begin{tabular}{ccccccccc}
\hline
Model        &  ejecta & ejecta   & shock & $\nu\rs{max}$$^a$ & clump &
injection &  $\min(\gamma\rs{eff})$ & initial \\
abbreviation & profile & clump. & mod. &  & size$^b$ &
efficiency &  & age [yr] \\
\hline
REF-EX       & EXP$^c$ & no  & no & $-$ & $-$  & $-$ & $5/3$ & 10 \\
REF-PL       & PLAW$^d$   & no  & no & $-$ & $-$  & $-$ & $5/3$ & 10 \\
EX-C1.5-D1   & EXP & yes & no & 1.5 & 1\% & $-$ & $5/3$ & 10 \\
EX-C1.5-D2   & EXP & yes & no & 1.5 & 2\% & $-$ & $5/3$ & 10 \\
EX-C2.5-D1   & EXP & yes & no & 2.5 & 1\% & $-$ & $5/3$ & 10 \\
EX-C2.5-D2   & EXP & yes & no & 2.5 & 2\% & $-$ & $5/3$ & 10 \\
EX-C5.0-D1   & EXP & yes & no & 5.0 & 1\% & $-$ & $5/3$ & 10 \\
EX-C5.0-D2   & EXP & yes & no & 5.0 & 2\% & $-$ & $5/3$ & 10 \\
PL-C1.5-D1   & PLAW   & yes & no & 1.5 & 1\% & $-$ & $5/3$ & 10 \\
PL-C1.5-D2   & PLAW   & yes & no & 1.5 & 2\% & $-$ & $5/3$ & 10 \\
PL-C2.5-D1   & PLAW   & yes & no & 2.5 & 1\% & $-$ & $5/3$ & 10 \\
PL-C2.5-D2   & PLAW   & yes & no & 2.5 & 2\% & $-$ & $5/3$ & 10 \\
PL-C5.0-D1   & PLAW   & yes & no & 5.0 & 1\% & $-$ & $5/3$ & 10 \\
PL-C5.0-D2   & PLAW   & yes & no & 5.0 & 2\% & $-$ & $5/3$ & 10 \\
EX-QPAR-G1.1 & EXP & no & yes & $-$ & $-$  & QPAR$^e$ & $1.1$ & 10 \\
EX-QPAR-G1.3 & EXP & no & yes & $-$ & $-$  & QPAR & $4/3$ & 10 \\
PL-QPAR-G1.1 & PLAW   & no & yes & $-$ & $-$  & QPAR & $1.1$ & 10 \\
PL-QPAR-G1.3 & PLAW   & no & yes & $-$ & $-$  & QPAR & $4/3$ & 10 \\
EX-ISO-G1.1  & EXP & no & yes & $-$ & $-$  & ISO$^f$ & $1.1$ & 10 \\
PL-ISO-G1.1  & PLAW   & no & yes & $-$ & $-$  & ISO & $1.1$ & 10 \\
EX-C3.5-D1-QPAR-G1.3  & EXP & yes & yes & 3.5 & 1\% & QPAR & $4/3$ & 10 \\
EX-C3.5-D2-QPAR-G1.3  & EXP & yes & yes & 3.5 & 2\% & QPAR & $4/3$ & 10 \\
PL-C3.5-D1-QPAR-G1.3  & PLAW & yes & yes & 3.5 & 1\% & QPAR & $4/3$ & 10 \\
PL-C3.5-D2-QPAR-G1.3  & PLAW & yes & yes & 3.5 & 2\% & QPAR & $4/3$ & 10 \\
EX-C1.5-D1-2YR   & EXP & yes & no & 1.5 & 1\% & $-$ & $5/3$ & 5 \\
EX-C5.0-D2-2YR   & EXP & yes & no & 5.0 & 2\% & $-$ & $5/3$ & 5 \\
\hline
\end{tabular}
\end{center}
$^a$ Maximum perturbation of mass density; 
$^b$ percentage of the initial diameter of the remnant;
$^c$ exponential profile;
$^d$ \referee{power-law} profile;
$^e$ quasi-parallel injection;
$^f$ isotropic injection.
\end{table*}%

Note that the ejecta clumps are presumably relics of the
deflagration of the outer layers of the exploding star (as suggested by
theoretical arguments and observations). In principle, therefore, the
clumps are expected to be concentrated in a shell within the ballistically
expanding ejecta rather than being distributed in the whole unshocked
ejecta as done here. On the other hand, in our simulations, the ramp
profile of the initial velocity of the ejecta makes the clumps in the
outer layers those with the highest speed, so that the shocked ISM is
mostly perturbed by such clumps. Concerning the focus of this paper,
namely the structure of the RT mixing in the region between the forward
and reverse shocks, we do not expect therefore significant changes to the
our results if considering a distribution of clumps concentrated in the
outer layers of the ejecta.

As initial conditions, we adopted parameters appropriate to reproducing
the SNR SN\,1006 after 1000~yr of evolution: we assumed an initial
spherical remnant with radius $R\rs{fs0} = 0.5$~pc (corresponding to
an initial age of $\approx 10$~yr), originating in a progenitor star
with mass of $1.4~M_{\rm sun}$, and propagating through an unperturbed
magneto-static medium. \referee{Note that we payed particular attention
to have an initial total mass of ejecta $M\rs{ej} = 1.4~M_{\rm sun}$
in all the simulations considered here, including those with a clumpy
structure of the ejecta.} The initial total energy $E_0 = 1.5\times
10^{51}$ ergs leads to a remnant radius $R\rs{snr} \approx 8.5$ pc
at $t=1000$~yr and is partitioned so that $>99$\% of the SN energy is
kinetic. The remnant expands through a homogeneous isothermal medium
of plasma number density $n = 0.05$ cm$^{-3}$ and temperature $T=10^4$
K. The initial ambient magnetic field configuration is that suggested
by \citet{2011A&A...531A.129B} for SN\,1006 and resulting from the
comparison of radio observations of SN\,1006 with MHD models: the ambient
magnetic field is characterized by a non-zero gradient of its strength
perpendicular to the average magnetic field that leads to a variation
of $|\vec{B}|$ of about a factor 1.4 over a scale of 10~pc. In all our
simulations, the magnetic field strength is $\approx 3\,\mu$G in the
environment of the explosion site. We follow the remnant evolution for
1000 yr.

The computational domain extends 24~pc in the $x$, $y$, and $z$
directions. Special emphasis was placed on capturing the enormous range in
spatial scales in the remnant. To this end, we exploited the adaptive mesh
capabilities of the \FLASH\ code by using 11 nested levels of resolution,
with resolution increasing twice at each refinement level. The
refinement/derefinement criterion adopted (\citealt{loehner}) follows the
changes in mass density, temperature, and tracer of ejecta. In addition,
the calculations were performed using also an automatic mesh derefinement
scheme in the whole spatial domain that kept the computational cost
approximately constant as the blast expanded: the maximum number of
refinement levels used in the calculation gradually decreased from 11
(initially) to 7 (at the final time) following the expansion of the blast
and keeping roughly the same number of grid zones per radius of the
remnant. At the beginning (at the end) of the simulation, this grid
configuration yielded an effective resolution of $\approx 2.9\times
10^{-3}$~pc ($\approx 4.6\times 10^{-2}$~pc) at the finest level,
corresponding to $\approx 170$ zones per initial radius of the remnant
($\approx 190$ zones per final radius of the remnant). The effective
mesh size varied from $8192^3$ initially to $512^3$ at the final time.

We also performed two additional simulations with the same parameters
of runs EX-C1.5-D1 and EX-C5.0-D2 but starting as early as $\approx
2$~yr after the SN explosion (the initial spherical remnant has radius
$R\rs{fs0} = 0.125$~pc) to check if the results depend on the time when
the clumpy structure of the ejecta is initialized. The results of this
comparison are discussed in Appendix~\ref{app1}.

\section{Results}
\label{sec3}

\subsection{Effects of back-reaction of accelerated cosmic rays}
\label{sec3.1}

As a first step, we analyzed the effects of back-reaction of
accelerated CRs on the separation between the blast wave and the
contact discontinuity, by considering models accounting for the shock
modification by accelerated CR particles but without initial clumping of
ejecta\footnote{Note that, in these simulations, we did not introduce
any seed perturbation. The departures from spherical symmetry are
entirely due to the mesh and to possible numerical fluctuations.}. A
recent comprehensive study of these effects on the development of RT
instabilities in young SNRs is given by \cite{2011MNRAS.415...83W} (see
also references therein). Our study differs from previous works in that
it includes magnetic fields and a possible dependence of the CR particle
acceleration on the obliquity angle. In particular, we focused on the
isotropic and quasi-parallel scenario discussed in Sect.~\ref{sec2}; the
results for models assuming quasi-perpendicular injection are expected
to be analogous to those discussed here for quasi-parallel injection,
showing a modulation of the shock modification with the obliquity angle.

\begin{figure}[!t]
  \centering
  \includegraphics[width=8.5cm]{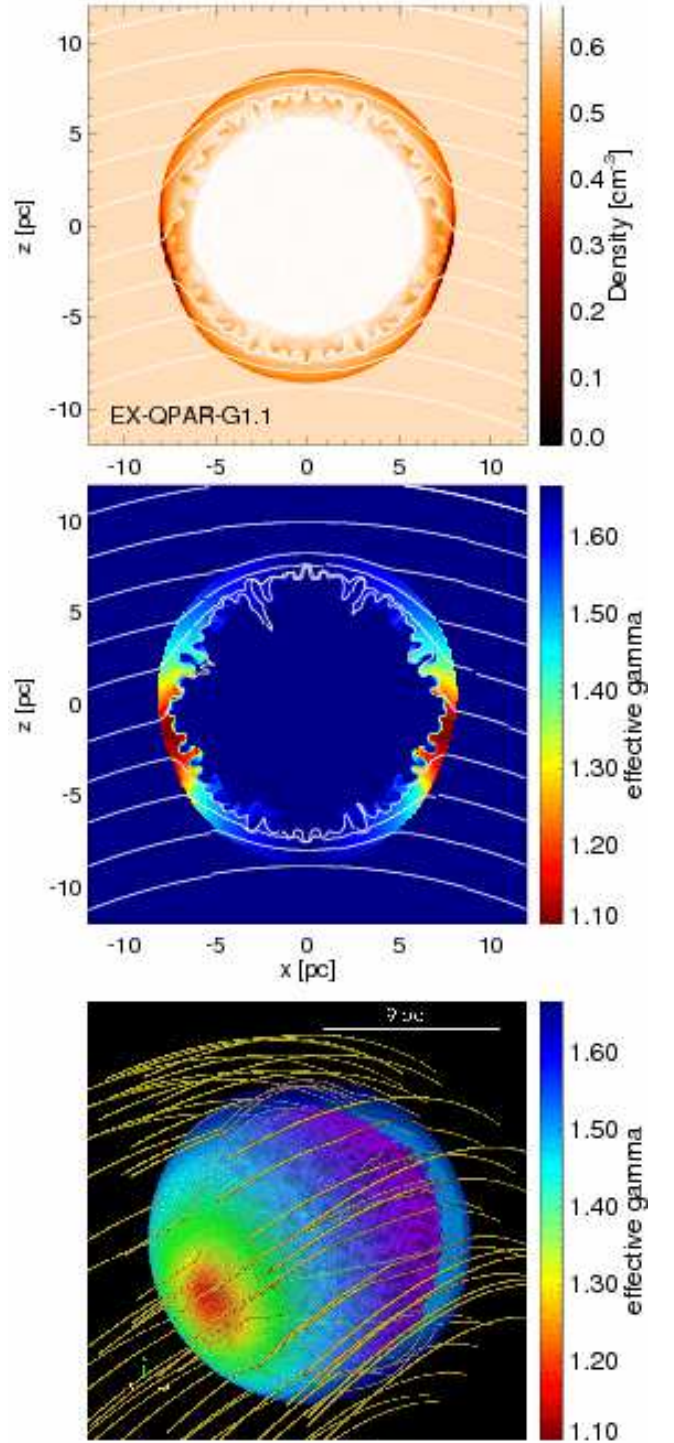}
  \caption{2D sections in the $(x,z)$ plane of the spatial distribution of
           plasma number density (upper panel) and effective adiabatic
           index (middle panel), after 1000~yr of evolution, for a
           model accounting for the shock modification by accelerated CRs
           but without initial clumping of ejecta (run EX-QPAR-G1.1). The
           lower panel shows the 3D volume rendering describing the
           spatial distribution of the effective adiabatic index. The
           index $\gamma_{\rm eff}$ is minimum in red regions (see
           colour bar). The white lines in the upper two panels and 
           the yellow lines in the lower panel are sampled magnetic 
           field lines. The violet surface in the lower panel tracks
           the ejecta material.}
  \label{fig3}
\end{figure}

As expected for cases in which the magnetic field has a
component parallel to the surface of the contact discontinuity
(\citealt{1961hhs..book.....C}), the magnetic field limits the growth
of hydrodynamic instabilities through the tension of field lines which
maintain a more laminar flow around the contact discontinuity. The energy
losses to CRs at the forward shock lead to a greater shock compression
ratio in all the cases examined (see also \citealt{2001ApJ...560..244B,
2011MNRAS.415...83W}). As a consequence, the density of the shocked ISM
is greater and the separation between the blast wave and the contact
discontinuity is shorter than predicted for a non-modified shock in
regions with $\gamma_{\rm eff} < 5/3$, i.e. where the back-reaction of
accelerated CRs is efficient. In the quasi-parallel case, since the
back-reaction of CRs is more effective at parallel shocks, the shock
modification is modulated with the obliquity angle. As an example
of this case, Fig.~\ref{fig3} presents the results for a model with
an exponential profile of the initial ejecta density after 1000~yr of
evolution (run EX-QPAR-G1.1; see Table~\ref{tab1}). In this model we also
assumed extreme energy losses to accelerate the CRs, so that the minimum
effective adiabatic index is $\gamma_{\rm eff} = 1.1$. The modulation
of the back-reaction of accelerated CRs with the obliquity angle is
evident in the figure, showing a larger compressibility and higher
values of plasma density at parallel shocks. Such a modulation is absent
in the isotropic case where the effects of CR particle acceleration
are the same everywhere at the shock front (runs EX-ISO-G1.1 and
PL-ISO-G1.1). In these cases the plasma compressibility is the
largest everywhere at the shock front, and the post-shock magnetic field
can reach values up to $\approx 50-70\,\mu$G at perpendicular shocks. It
is worth mentioning that in both the quasi-parallel and isotropic cases,
the simulations do not show any significant perturbation of the remnant
outline and occurrence of protrusions after 1000~yr of evolution, even
assuming extreme energy losses to accelerate the CRs. These results are
in agreement with previous studies showing that enhanced RT mixing due
to efficient particle acceleration determines only a slight perturbation
of the forward shock near the epoch of young SNRs as SN\,1006 or Tycho
(e.g. \citealt{2001ApJ...560..244B, 2011MNRAS.415...83W}).

We investigated the effect of accelerated CRs on the separation
between the blast wave and the contact discontinuity, by deriving the
azimuthal profiles of the ratio of the forward shock radius to the
contact discontinuity radius $R\rs{fs}/R\rs{cd}$ from the models. The
position of the forward shock was estimated from 2D maps of projected
emission-measure-weighted temperature $\langle T\rangle$ as the jump
in $\langle T\rangle$ in the direction of compression (determined by
looking at the velocity field) at temperatures $T > 1$~MK. The position
of the contact discontinuity was estimated by using the passive tracer
$C\rs{ej}$ included in the model (see Sect.~\ref{sec2}): during the
remnant evolution, the ejecta and the shocked ISM mix together, leading
to regions with $0 < C\rs{ej} < 1$; at any time $t$ the density of ejecta
material in a fluid cell is given by $\rho\rs{ej} = \rho C\rs{ej}$. We
derived the position of the contact discontinuity from 2D maps of
projected $\rho\rs{ej}$ as the local peak of $\rho\rs{ej}$ closest
to the forward shock in the direction of compression. The azimuthal
profiles derived from the models in such a way are directly comparable
with observations and, in particular, with the profiles derived by
\cite{2009A&A...501..239M} in the analysis of the observations of
SN\,1006 (see Fig.~6 in \citealt{2009A&A...501..239M}).

\begin{figure}[!t]
  \centering
  \includegraphics[width=8.5cm]{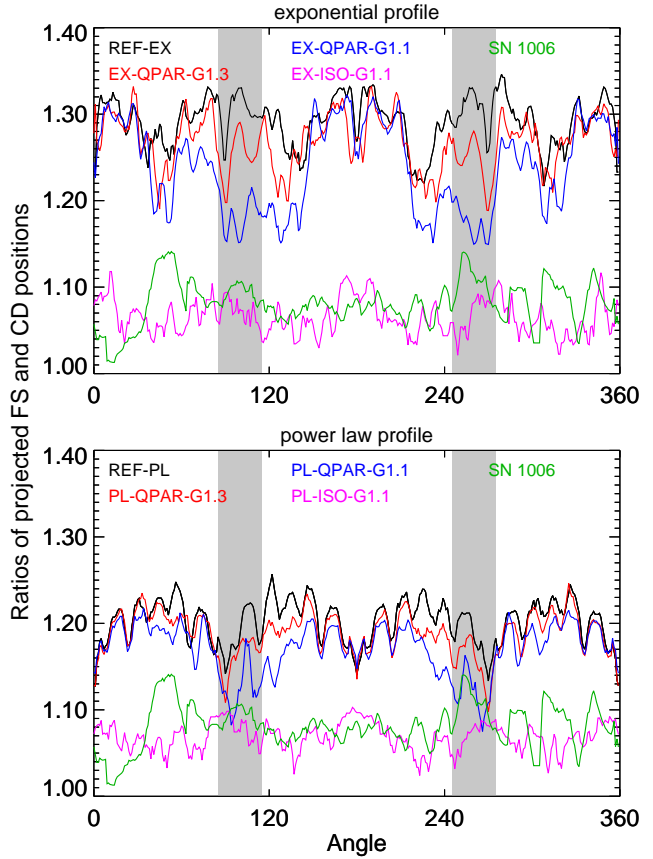}
  \caption{Azimuthal profiles of the ratio of the forward shock radius
           to the contact discontinuity radius $R\rs{fs}/R\rs{cd}$ for
           models without ejecta clumping and with initial ejecta density
           profile either exponential (upper panel) or \referee{power-law}
           (lower panel). The black lines mark the profiles derived from
           the reference models not accounting for the back-reaction of
           accelerated CRs (runs REF-EX and REF-PL). The red and blue
           lines mark the profiles derived from models including the
           shock modification modulated by the obliquity angle either with
           minimum $\gamma\rs{eff} = 4/3$ (red lines; runs EX-QPAR-G1.3
           and PL-QPAR-G1.3) or minimum $\gamma\rs{eff} = 1.1$ (blue
           lines; runs EX-QPAR-G1.1 and PL-QPAR-G1.1). The magenta lines
           mark the profiles derived from models including the shock
           modification with no obliquity dependence and $\gamma\rs{eff}
           = 1.1$ (runs EX-ISO-G1.1 and PL-ISO-G1.1).  The green line
           marks the profile derived from the observations of SN\,1006
           (\citealt{2009A&A...501..239M}). The gray stripes
           mark the regions where the acceleration of CRs is the largest.}
  \label{fig4}
\end{figure}

Fig.~\ref{fig4} shows the azimuthal profiles of $R\rs{fs}/R\rs{cd}$
derived from the models when the aspect angle is $90^0$ (i.e when the
average magnetic field is perpendicular to the line of sight). The
green line is the profile derived from the observations of SN\,1006
(\citealt{2009A&A...501..239M}) and the black lines are the reference
models with no shock modification and no ejecta clumping. The gray
stripes mark the regions where, in the models, the acceleration of CRs
is the largest. In the quasi-parallel case, we found that the modeled
profiles are modulated by the obliquity angle and, in general, are higher
than those observed (see red and blue lines in Fig.~\ref{fig4}). The
observations can be reproduced only in limited regions where the effect
of accelerated CRs is the highest. Aspect angles lower than $90^0$
make the comparison between models and observations worse because the
regions of efficient CRs acceleration would not be at the limb (so that
the ratio $R\rs{fs}/R\rs{cd}$ increases). On the other hand, the models
with no obliquity dependence of the back-reaction of accelerated CRs
and $\gamma\rs{eff} = 1.1$ reproduce the observed profiles quite well
(magenta lines in Fig.~\ref{fig4}). These results suggest that the
observations could be reproduced only if the back-reaction of accelerated
CRs is extreme (i.e. $\gamma\rs{eff} \approx 1.1$) and independent of
obliquity angle (i.e. the CRs acceleration and escape should be ubiquitous
at the forward shock).

\subsection{Effects of ejecta clumping and instability}
\label{sec:clump}

As a next step, we investigated the effects of ejecta clumping on
the evolution and morphology of the remnant by considering models
without back-reaction of accelerated CRs and accounting only for the
ejecta clumping. In addition to the spectropolarimetric studies of SNe
discussed in Sect.~\ref{sec1} (see also \citealt{2010ApJ...720.1500H} and
references therein), a widespread clumpiness of ejecta is also suggested
by X-ray and radio observations, showing knots located near the edge
of the remnants, and outward protrusions in many cases surrounding
the knots (e.g. \citealt{1997ApJ...475..665H, 1998A&A...334.1060V,
2011ApJ...735L..21R}). All these features cannot be explained by
instabilities generated by linear perturbations and have been interpreted
as being due to clumps of ejecta expanding into the intershock region
(e.g. \citealt{2001ApJ...549.1119W}). The interactions among the clumps
of ejecta are expected to contribute to seed the RT instabilities and
enhance their growth, thus strongly influencing the final morphology of
the remnant.

The basic physics of the evolution of a single clump of ejecta
expanding through the intershock structure of a SNR is similar
to that for the interaction of a shock with a cloud of the ISM
(e.g. \citealt{1994ApJ...420..213K}) and has been extensively discussed
by \citet{2001ApJ...549.1119W}. The major factors in the clump-remnant
interaction are the density contrast of the clump with respect to the ISM,
the clump size, and the position of the clump in the initial distribution
of ejecta (or, alternatively, the time of initiation of the clump-shock
interaction). In general, after passing through the reverse shock, the
single clump evolves toward a core-plume structure with a crescent-like
shape characterized by Kelvin-Helmholtz (KH) instabilities developing
in the downstream region. As the clump travels through the intershock
structure, RT instabilities develops on the upstream side of the clump,
leading to its progressive fragmentation. Depending on its initial
density contrast, size, and time of initiation of the clump-shock
interaction, the clump can reach the forward shock, causing a bulge
on the remnant outline as the ram pressure pushes material ahead (see
\citealt{2001ApJ...549.1119W} for a detailed description). This is the
way ejecta protrusions form. After the clump is completely fragmented,
the bulge (the protrusion) disappears and the clump fragments are mixed
with the shocked ISM and swept back in the remnant. The perturbation of
the forward shock front by the interaction with the clumps is more likely
during the early phases of the remnant evolution when the density contrast
between the ejecta clumps and the ISM is larger.

In our case, we are assuming that the ejecta structure is formed by
hundreds of thousands of clumps modelled as per-cell random density
perturbations (see Sect.~\ref{sec2} and Fig.~\ref{fig2}); in each
simulation, the clumps have the same size and are characterized by
different density contrasts (i.e. different density perturbations)
and different positions (i.e. the time of initiation of the interaction
of each clump with the reverse shock is different). The clump-remnant
interaction therefore is complicated by the multiple interactions among
clumps with different density contrast and velocity. In addition,
our model includes the magnetic field which is known to limit the
growth of hydrodynamic instabilities in the shock-cloud interaction
(e.g. \citealt{1994ApJ...433..757M, 1996ApJ...473..365J}) due to the
tension of the magnetic field lines which maintain a more laminar
flow around the cloud surface (see also \citealt{2005ApJ...619..327F,
2008ApJ...678..274O}). In the present case, during the clump evolution,
the magnetic field is expected to be trapped at the nose of the clump,
leading to a continuous increase of the magnetic pressure and field
tension there that limit the growth of RT instabilities responsible for
the clump fragmentation. As a result, the clumps are expected to survive
for a longer time than those studied by \citet{2001ApJ...549.1119W}
(their simulations do not include the magnetic field), increasing
their probability to reach the forward shock.

\begin{figure}[!t]
  \centering
  \includegraphics[width=8.5cm]{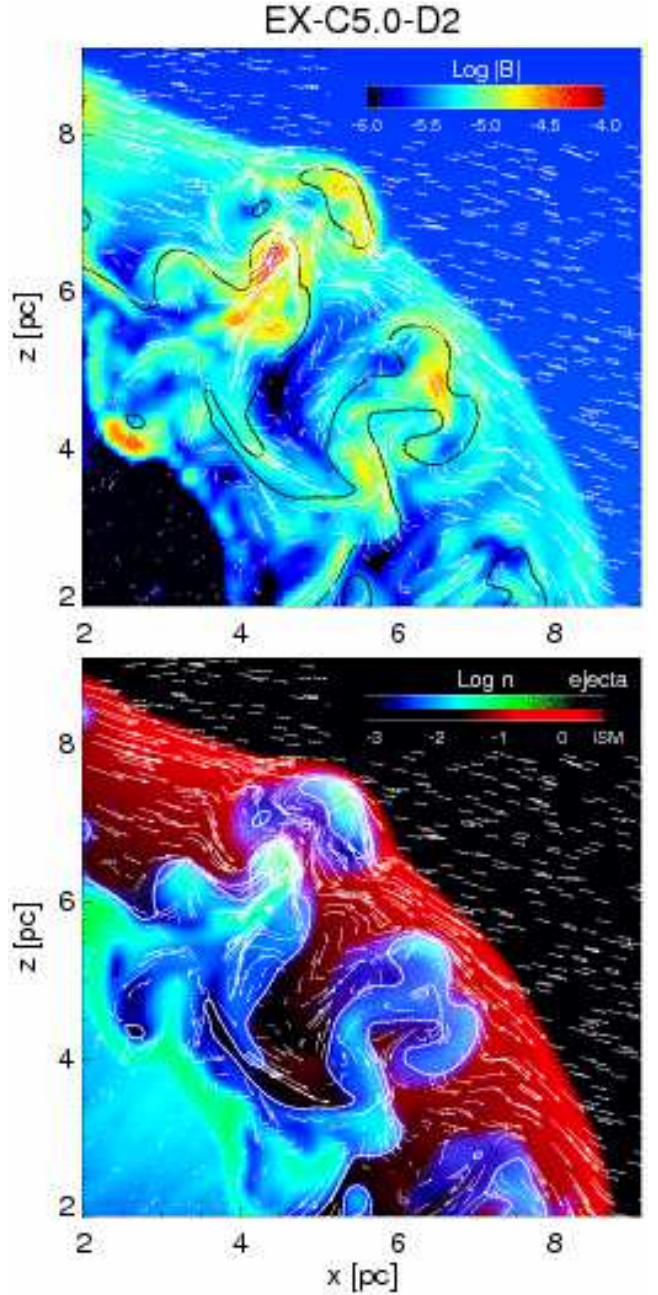}
  \caption{Close-up view of the remnant limb for the model EX-C5.0-D2 at
  $t=1000$~yr, showing a colour-coded cross-section image of the magnetic
  field strength (G; upper panel) and a composite cross-section image
  (lower panel) combining the plasma density (cm$^{-3}$) of the shocked
  ISM (red) and that of the ejecta (blue-green). The contours enclose
  the cells consisting of the original ejecta material by more than
  10\%. The magnetic field is described by the superimposed arrows the
  length of which is proportional to the magnitude of the field vector.}
  \label{fig5}
\end{figure}

\begin{figure*}[!th]
  \centering
  \includegraphics[width=16cm]{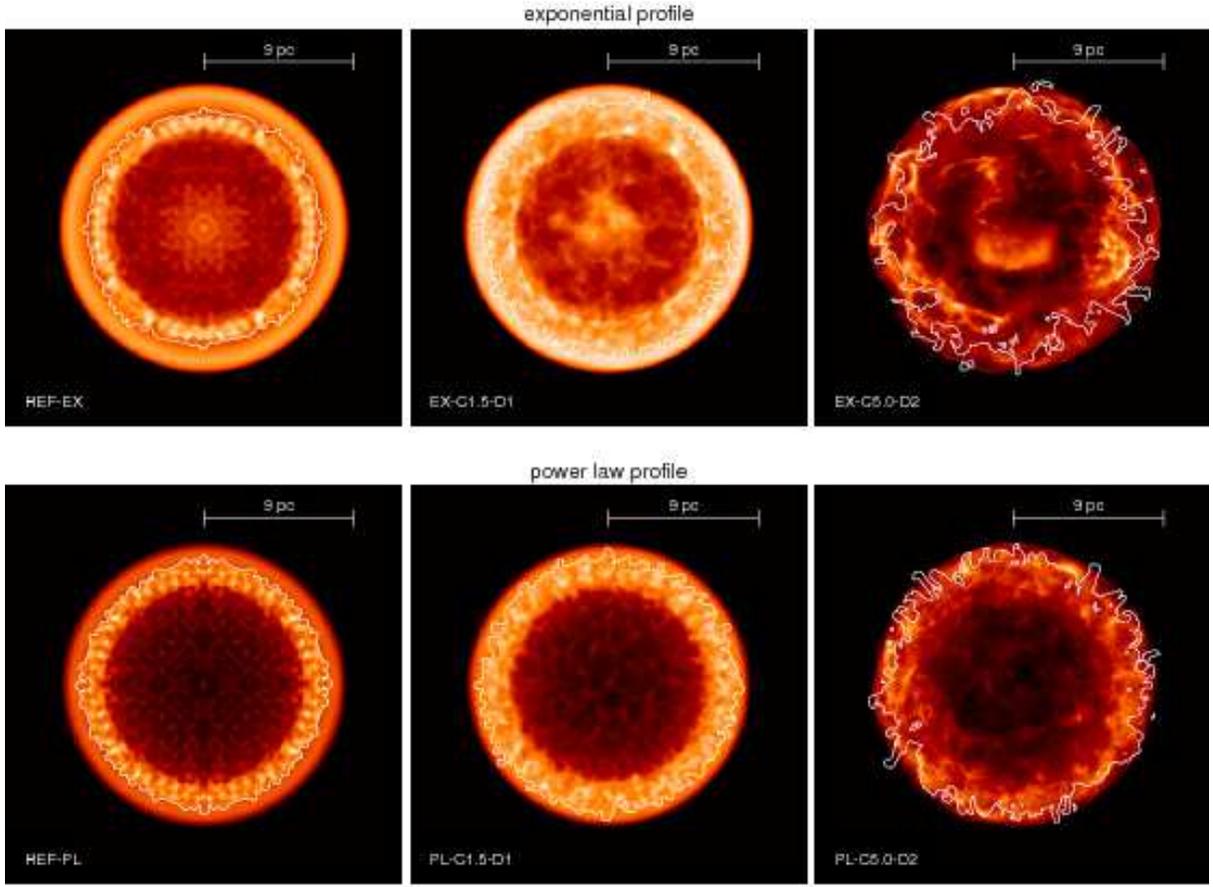}
  \caption{3D rendering of mass density for models with an exponential
          (upper panels) or a \referee{power-law} (lower panels) initial profile
          of ejecta and no shock modification by CRs, after 1000~yr of
          evolution. Left panels show the maps derived from the reference
          models not including the ejecta clumping (runs REF-EX and
          REF-PL). The initial clumpy structure of the ejecta is
          characterized by clumps either with size $\approx 1$\% of the
          initial diameter of the remnant $D\rs{snr0}$ and maximum
          density perturbation $\nu\rs{max} = 1.5$ (center panels)
          or with size $\approx 2$\% of the initial diameter of the
          remnant and maximum density perturbation $\nu\rs{max} = 5.0$
          (right panels). The white contours enclose the ejecta material
          (i.e. cells consisting of the original ejecta material by more
          than 50\%).}
  \label{fig6}
\end{figure*}

As an example, Fig.~\ref{fig5} shows a close-up view of the
remnant limb for the model EX-C5.0-D2, illustrating the magnetic
field strength (upper panel) and the plasma density distribution
(lower panel) at $t=1000$~yr. As expected, the magnetic field follows
the plasma structures formed during the evolution of the clumps with
preferentially radial components around the RT fingers. The magnetic
field is strongly modified by the clumps and it can be enhanced by up
to two orders of magnitude ($|\vec{B}|\approx 100\,\mu$G, whereas the
unperturbed magnetic field strength is $\approx 2.5\,\mu$G) in the ejecta
clumps (see red regions in the upper panel of Fig.~\ref{fig5}). Note that,
in model EX-C5.0-D2, no back-reaction of accelerated CRs is taken into
account and the magnetic field in inter-clumps regions at the forward
shock ($|\vec{B}|\approx 10\,\mu$G) is that predicted for non-modified
shocks, namely much lower than that measured in the X-ray rims of SN\,1006
($50 \lsim |\vec{B}| \lsim 150\,\mu$G; \citealt{2003A&A...412L..11B,
2009A&A...505..169B, 2010A&A...516A..62A, 2011MNRAS.413.1643P,
2012MNRAS.419..608P}).

As examples, Fig.~\ref{fig6} shows the 3D rendering of plasma density for
the reference cases without clumping (runs REF-EX and REF-PL) and for
the limit cases with clumping considered in this paper, namely models
with an ejecta structure characterized either by clumps with small
size and low density perturbations (runs EX-C1.5-D1 and PL-C1.5-D1
in Table~\ref{tab1}) or by clumps with large size and high density
perturbations (runs EX-C5.0-D2, and PL-C5.0-D2). The figure shows that
the enhanced intershock RT mixing can easily spread the ejecta material
close to, or even beyond, the average radius of the forward shock,
depending on the size and density contrast of the initial clumps. This
can occur very soon after the explosion, depending again on the size and
density contrast of the clumps seeding the instabilities. As a result,
we found that: 1) the RT mixing reaches the forward shock front possibly
perturbing the remnant outline, 2) knots and filamentary structures
characterize the remnant morphology, and 3) clumps of ejecta can be
very close to or even protruding beyond the main blast wave leading
to evident knots near the remnant edge as observed, for instance,
in SN\,1006 (\citealt{2011ApJ...735L..21R}) and Tycho's SNR (e.g.
\citealt{1998A&A...334.1060V}). In general, increasing the initial size
of the clumps or their density perturbation (i.e. going from the left to
the right panel of Fig.~\ref{fig6}), both the perturbation of the remnant
outline and the occurrence of ejecta protrusions increase. Fig.~\ref{fig6}
also shows that the characteristic size of the density features formed
within the remnant is comparable to that of the features observed,
for instance, in SN\,1006.

\begin{figure}[!t]
  \centering
  \includegraphics[width=8cm]{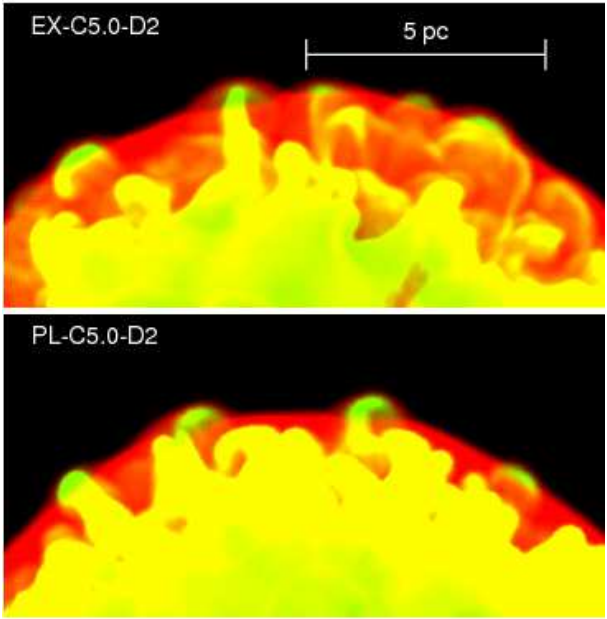}
  \caption{Composite images of the remnant limb showing the square
           of plasma density of the shocked ISM (red) and that of
           the ejecta (green and yellow), both projected along the
           line-of-sight, for models EX-C5.0-D2 (upper panel) and
           PL-C5.0-D2 (lower panel) after 1000~yr of evolution.}
  \label{fig7}
\end{figure}

A remarkable feature of the simulations including the ejecta clumping is
the occurrence of several protrusions due to clumps of ejecta overtaking
the forward shock. Fig.~\ref{fig7} shows composite images of the SNR
combining the square of plasma density of the shocked ISM (red) and that
of the ejecta (green and yellow), both projected along the line-of-sight,
for models EX-C5.0-D2 and PL-C5.0-D2. The protrusions are evident
in both cases and are due to clumps with high density contrast
originating from the outer layers of the ejecta. Our calculations show
that the number of protrusions at $t=1000$~yr is higher for larger size
of the clumps and higher density contrasts of the clumps and decreases
with the age of the remnant. In fact the simulations showed that,
during the remnant evolution, new protrusions are continuously formed
and, subsequently, disappear when the clumps responsible for them are
decelerated and the forward shock front catches up with them (see also
\citealt{2002ApJ...574..155W}). In this process, the clumps contribute
in the perturbation of the remnant outline and in the formation of plasma
features in the outer part of the remnant.

\begin{figure}[!t]
  \centering
  \includegraphics[width=8.5cm]{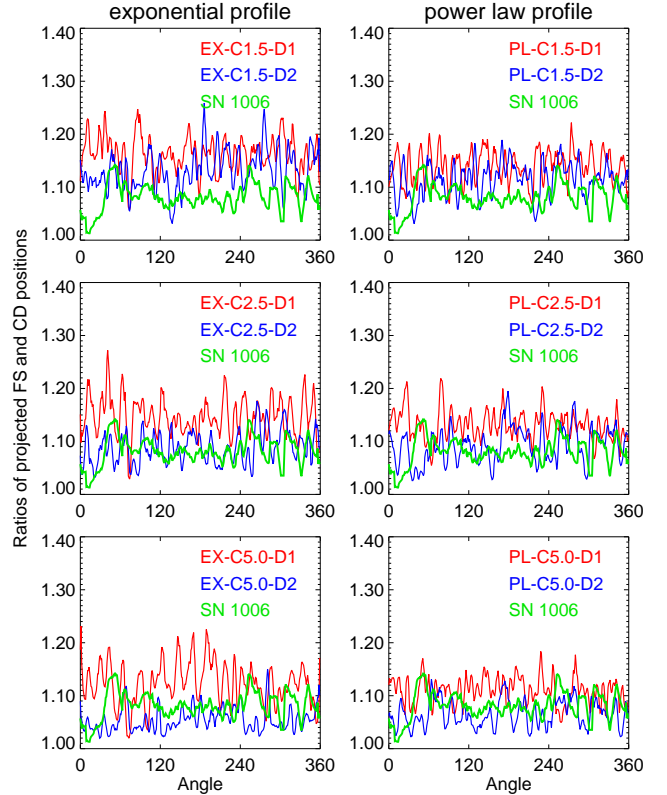}
  \caption{Azimuthal profiles of the ratio of the forward shock radius
           to the contact discontinuity radius $R\rs{fs}/R\rs{cd}$ for
           models without back-reaction of accelerated CRs and with
           ejecta clumping and initial ejecta density profile either
           exponential (left panels) or \referee{power-law} (right panels). The
           figure shows the profiles derived from models with maximum
           density perturbation $\nu\rs{max} = 1.5$ (upper panels),
           $\nu\rs{max} = 2.5$ (middle), and $\nu\rs{max} = 5.0$ (lower),
           and with initial size of the clumps $\approx 1$\% (red lines)
           and $\approx 2$\% (blue lines) of the initial diameter of
           the remnant $D\rs{snr0}$. The green line marks the
           profile derived from the observations of SN\,1006
           (\citealt{2009A&A...501..239M}).}
  \label{fig8}
\end{figure}

Also in this case, we compared the azimuthal profiles of the ratio
$R\rs{fs}/R\rs{cd}$ derived from the models with that observed in SN\,1006
(see Fig.~\ref{fig8}). We found that the initial clumping of ejecta
makes the azimuthal profiles of $R\rs{fs}/R\rs{cd}$ fairly uniform
and lower than expected for models without a clumpy structure of the
ejecta and comparable with models accounting for extreme and ubiquitous
acceleration of CR particles at the forward shock (i.e. isotropic
models with $\gamma\rs{eff}\approx 1.1$; compare Fig.~\ref{fig4}
and Fig.~\ref{fig8}). In particular, we found that, in the case of
SN\,1006, the observed profile can be reproduced by models with a maximum
density perturbation of ejecta $\nu\rs{max}$ ranging between 2.5 and 5,
and with initial size of ejecta clumps of the order of $2$\%
of the initial diameter of the remnant (see the blue lines in middle and
lower panels in Fig.~\ref{fig8}).

\subsection{Ejecta clumping and cosmic rays acceleration}

\begin{figure}[!t]
  \centering
  \includegraphics[width=8.5cm]{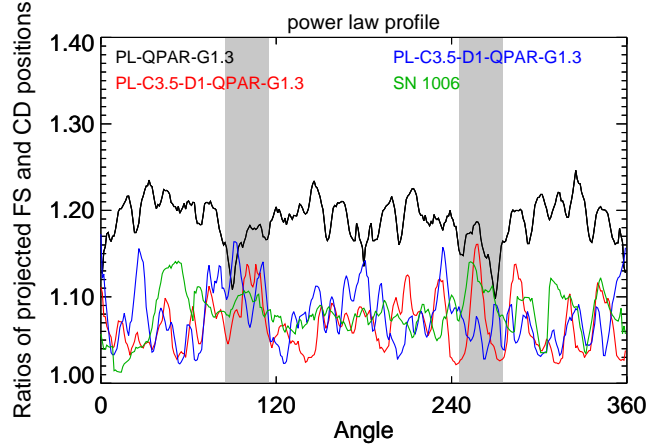}
  \caption{As in Fig.~\ref{fig4} for a model accounting for the
           back-reaction of accelerated CRs only (run PL-QPAR-G1.3)
           and for models including both the CR particle acceleration
           and the ejecta clumping (runs PL-C3.5-D1-QPAR-G1.3,
           and PL-C3.5-D2-QPAR-G1.3). In all the cases, the
           minimum $\gamma\rs{eff} = 4/3$. The green line marks
           the profile derived from the observations of SN\,1006
           (\citealt{2009A&A...501..239M}).  The gray stripes mark the
           regions where the acceleration of CRs is the largest.}
  \label{fig9}
\end{figure}

As a last step, we have investigated the effects of back-reaction of
accelerated CRs on the remnant morphology in the presence of ejecta
clumping through simulations including both physical processes (runs
EX-C3.5-D1-QPAR-G1.3, EX-C3.5-D2-QPAR-G1.3, PL-C3.5-D1-QPAR-G1.3,
and PL-C3.5-D2-QPAR-G1.3 in Table~\ref{tab1}). We found that
when the CR acceleration efficiency depends on the obliquity angle
(e.g. quasi-parallel models), the modulation of the shock modification
with the obliquity angle is not appreciable in the presence of ejecta
clumping (see Fig.~\ref{fig9}). In other words, our model predict that
the ejecta clumping can wash out the CR back-reaction effects on the
separation between the forward shock and the contact discontinuity. On the
other hand, our simulations have shown that the effects of back-reaction
of accelerated CRs can still be visible on the azimuthal profile of
plasma density which shows local maxima where the acceleration of CRs
is the largest (the plasma compressibility being the highest there).

\begin{figure}[!t]
  \centering
  \includegraphics[width=8.5cm]{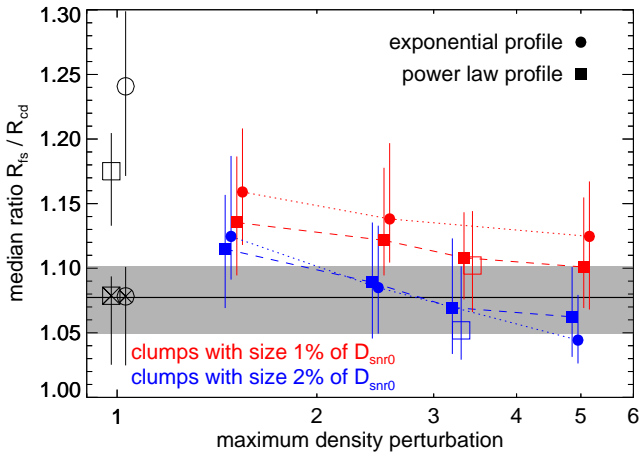}
  \caption{Median values of the ratio of the forward shock radius to the contact
           discontinuity radius $R\rs{fs}/R\rs{cd}$ versus the maximum
           density perturbation $\nu\rs{max}$ for models either with
           (red and blue symbols for initial size of the clumps $\approx
           1$\% and $\approx 2$\% of the remnant diameter $D\rs{snr0}$,
           respectively) or without (black symbols) ejecta clumping,
           and for models including the back-reaction of accelerated CRs
           (empty and crossed symbols for the quasi-parallel and isotropic
           cases, respectively). The grey region marks the range of values
           observed in SN\,1006 (\citealt{2009A&A...501..239M}).}
  \label{fig10}
\end{figure}

To make a more quantitative comparison between the model results and
the observations, we derived the median values of $R\rs{fs}/R\rs{cd}$
for each of the models in Table~\ref{tab1} and for the observed
profile. Fig.~\ref{fig10} shows the median values of $R\rs{fs}/R\rs{cd}$
versus the maximum density perturbation $\nu\rs{max}$ for models
accounting for only one of the effects considered in this paper (either
back-reaction of accelerated CRs or ejecta clumping) and for models
including both physical effects. We found that: the larger the size of
initial clumps of ejecta, the lower the value of the median ratio; the
higher the initial density perturbation, the lower the value of the median
ratio. The back-reaction of accelerated CRs slightly reduces the value
of the ratio in models accounting for the clumpy structure of the ejecta
(empty symbols in Fig.~\ref{fig10}) unless the energy losses to CRs are
large with an effective adiabatic index $\approx 1.1$ and ubiquitous at
the forward shock (as in the isotropic injection, see models EX-ISO-G1.1
and PL-ISO-G1.1; crossed symbols in Fig.~\ref{fig10}).

\section{Summary and conclusions}
\label{sec5}

We investigated the role of ejecta clumping and back-reaction of
accelerated CRs on the evolution and morphology of young Type Ia SNRs
and, in particular, on determining the observed separation between the
forward shock and the contact discontinuity and the high occurrence of
protrusions. To this end, we developed a 3D MHD model describing the
expansion of the remnant through a medium with nonuniform interstellar
magnetic field, including consistently the back-reaction of accelerated
CRs and the initial clumpy structure of the ejecta. We explored two
complementary cases in which one or the other of these physical processes
is turned either on or off in order to identify its effects on the remnant
evolution and morphology. Then we compared the model results with the
observations of SN\,1006 (\citealt{2009A&A...501..239M}). Particular
attention has been devoted to perform simulations with sufficient spatial
resolution to capture the details of the evolution of the clumps of
ejecta, exploiting the adaptive mesh refinement capabilities of the
\FLASH\ code.

As expected, we found that the acceleration of CR particles
makes the shell of shocked ISM thinner at the forward shock, thus
reducing the separation between the forward shock and the contact
discontinuity. Any dependence of the back-reaction of accelerated
CRs on the obliquity angle should be evident as a modulation of
the azimuthal profile of the ratio of the forward shock radius
to the contact discontinuity radius $R\rs{fs}/R\rs{cd}$. In the
case of SN\,1006, the comparison of the modelled profiles with
those observed shows that the back-reaction of accelerated CRs may
reproduce the observations only if the energy losses to CRs are extreme
(i.e. the effective adiabatic index is $\gamma\rs{eff}\approx 1.1$) and
independent of the obliquity angle (i.e. the effects of CR acceleration
are ubiquitous at the forward shock). In addition, the simulations have
shown that the large compression ratio due to the acceleration of CR
particles has no significant effect on the growth of RT instabilities,
in agreement with previous studies (e.g. \citealt{2001ApJ...560..244B,
2010A&A...515A.104F, 2011MNRAS.415...83W}). As a result, the remnant
outline is only slightly perturbed by the instabilities with very few
(if any) occurrence of protrusions near the epoch of young SNRs as
SN\,1006 or Tycho's SNR, even with very efficient acceleration of
CRs (see also \citealt{2011MNRAS.415...83W}). This fact contrasts
with the evidence of several protrusions observed in SN\,1006
(e.g. \citealt{2011ApJ...735L..21R}) and Tycho's SNR
(e.g. \citealt{1997ApJ...475..665H, 1998A&A...334.1060V}).

On the other hand, the clumpy structure of the ejecta can have important
consequences on the structure of the intershock RT mixing and on the final
morphology of the remnant. In particular, we found that the ejecta clumps
with the higher density contrasts approaching the contact discontinuity
enhance the growth of RT instabilities; RT fingers can easily reach
the forward shock and ejecta clumps can be found very close to,
or even beyond, the average shock radius with no need to invoke any
CR back-reaction at all to explain this phenomenon. As a result, the
separation between the forward shock and the contact discontinuity can
be significantly reduced, depending on the size and density contrast of
the clumps. In particular, we found that the larger the size of initial
clumps of ejecta and/or the higher their density contrast, the shorter the
width of the interaction region between the forward shock and the contact
discontinuity. The modelled azimuthal profile of $R\rs{fs}/R\rs{cd}$
is fairly uniform as observed in SN\,1006; the comparison of the model
results with the observations of SN\,1006 showed that the observed
profile of $R\rs{fs}/R\rs{cd}$ can be reproduced by models with a
maximum density perturbation of ejecta $\nu\rs{max}$ ranging between
2.5 and 5, and with initial size of ejecta clumps of the order of $2$\%
of the initial diameter of the remnant. We also found that the remnant
outline can be significantly perturbed by the enhanced RT fingers and,
in case of high density contrasts and large size of the clumps, several
protrusions can characterize the morphology of the remnant at the age
of SN\,1006.  Our study supports the idea that enhanced RT mixing due
to ejecta clumping can be responsible for the filamentary structures
and bumps seen on the outlines of young SNRs as SN\,1006 and Tycho's SNR.

Finally, our analysis has shown that the ejecta clumping, if present,
may wash out the effects of back-reaction of accelerated CRs on the
separation between the forward shock and the contact discontinuity. In
particular, if the CR acceleration efficiency depends on the obliquity
angle as, for instance, in the quasi-parallel scenario, the modulation of
the shock modification with the obliquity angle may be not appreciable in
the presence of ejecta clumping. We conclude therefore that, in general,
the separation between the forward shock and the contact discontinuity is
not a reliable diagnostic tool for studying the CR shock modification.

On the contrary, our model predicts that the effects of back-reaction
of accelerated CRs can still be appreciable on the azimuthal profile
of plasma density. In fact our simulations have shown that, even in the
presence of ejecta clumping, the density profile has local maxima where
the acceleration of CRs is the largest (the plasma compressibility
being the highest there). Also, due to the enhanced plasma
compressibility, the magnetic field strength can reach values of
$\approx 50-70\,\mu$G where the CR acceleration is the largest (see
Sect.~\ref{sec3.1}), that are comparable with those observed in the X-ray
rims of SN\,1006 (e.g. \citealt{2003A&A...412L..11B, 2009A&A...505..169B,
2010A&A...516A..62A, 2011MNRAS.413.1643P, 2012MNRAS.419..608P}). It
is interesting to note however that similar values of magnetic field
strength can also be reached locally in ejecta clumps close to the
forward shock with no need to invoke any CR back-reaction, but as a
result of the propagation of the clumps through the intershock region
(see Fig.~\ref{fig5}).

\bigskip
\acknowledgments
We thank an anonymous referee for useful suggestions.  This work was
supported in part by the Italian Ministry of University and Research
(MIUR) and by Istituto Nazionale di Astrofisica (INAF). The software used
in this work was in part developed by the DOE-supported ASC / Alliance
Center for Astrophysical Thermonuclear Flashes at the University
of Chicago. We acknowledge the CINECA Awards N. 998NZ7YK,2010 and
N. HP10C7MTR0,2011 for the availability of high performance computing
resources and support. Additional computations were carried out at the
SCAN\footnote{http://www.astropa.unipa.it/progetti\_ricerca/HPC/index.html}
(Sistema di Calcolo per l'Astrofisica Numerica) facility for high
performance computing at INAF -- Osservatorio Astronomico di Palermo. This
work was partially funded by the ASI-INAF contract n. I/009/10/0.

\bibliographystyle{apj}
\bibliography{references}

\newpage
\appendix
\section{Dependence of the results on the initial conditions}
\label{app1}

We checked the dependence of the results on the initial conditions and
in particular on the time when the clumpy structure of the ejecta is
initialized. To this end, we performed two additional simulations (runs
EX-C1.5-D1-2YR and EX-C5.0-D2-2YR in Table~\ref{tab1}) with the same
parameters of runs EX-C1.5-D1 and EX-C5.0-D2 but starting as early as
$\approx 2$~yr after the SN explosion (i.e. the initial spherical remnant
has radius $R\rs{fs0} = 0.125$~pc) instead of $\approx 10$~yr (with
$R\rs{fs0} = 0.5$~pc). In other words, we checked the dependence of the
results on the initial conditions for the limit cases considered in this
paper, namely models with a clumpy structure of the ejecta characterized
either by clumps with small size and low density perturbations (run
EX-C1.5-D1) or by clumps with large size and high density perturbations
(run EX-C5.0-D2). For runs EX-C1.5-D1-2YR and EX-C5.0-D2-2YR, we used 13
nested levels of resolution in the automatic mesh derefinement scheme to
keep the same spatial resolution as the other simulations discussed here
(i.e. $172$ zones per initial radius of the remnant); in this case the
effective mesh size was $32768\times 32768\times 32768$.

\begin{figure}[!t]
  \centering
  \includegraphics[width=8.5cm]{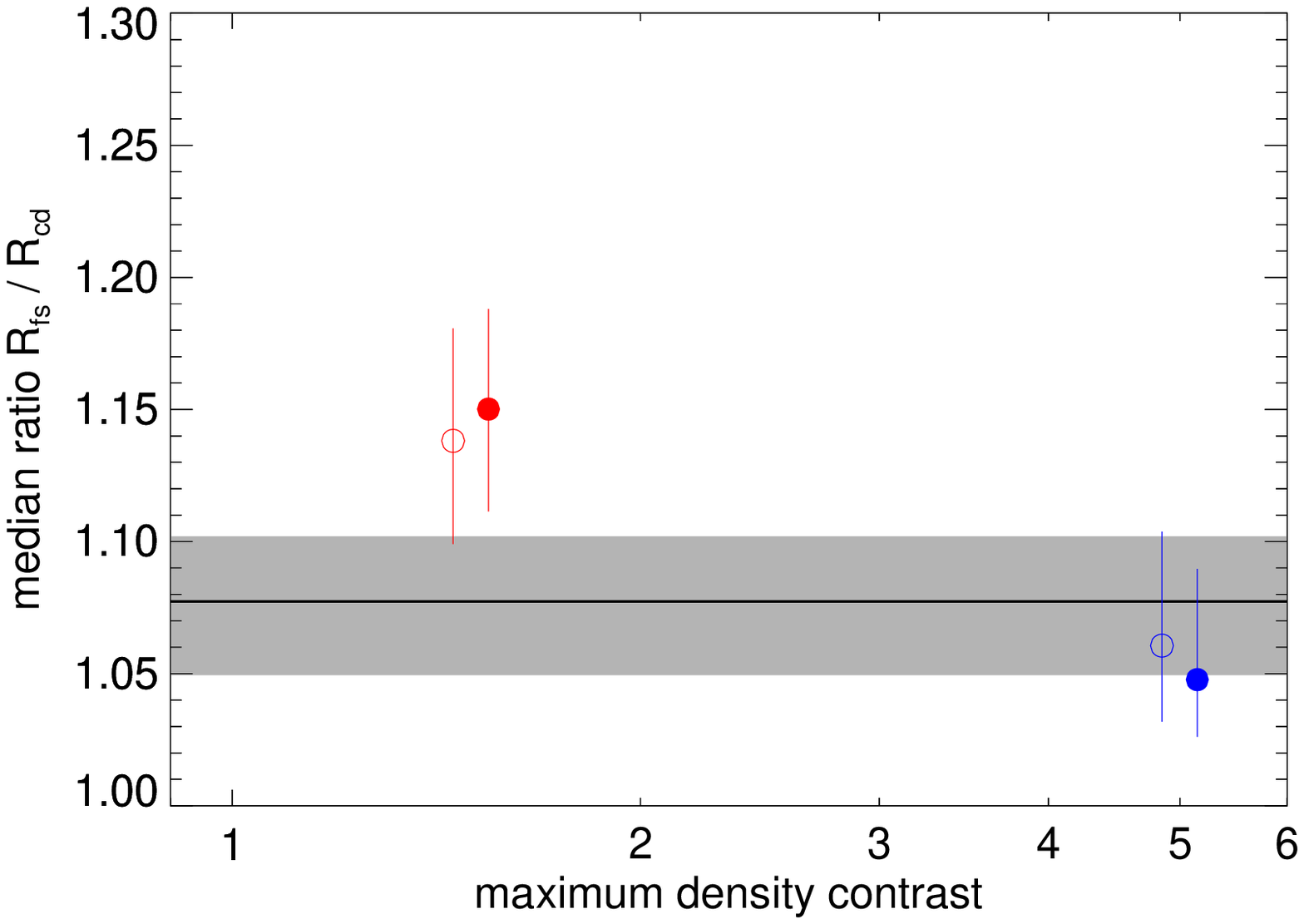}
  \caption{As in Fig.~\ref{fig10} for models starting as early as either
           $\approx 2$~yr (empty symbols; runs EX-C1.5-D1-2YR and
           EX-C5.0-D2-2YR) or $\approx 10$~yr (filled symbols; runs
           EX-C1.5-D1, EX-C5.0-D2) after the SN explosion. Red and
           blues symbols show the results for models with initial size
           of the clumps $\approx 1$\% and $\approx 2$\% of the remnant
           diameter $D\rs{snr0}$, respectively. The grey region marks
           the range of values
           observed in SN\,1006 (\citealt{2009A&A...501..239M}).}
  \label{fig_append}
\end{figure}

From the additional simulations, we derived the azimuthal profile of the
ratio of the forward shock radius to the contact discontinuity radius
$R\rs{fs}/R\rs{cd}$ as done for the other runs (see Fig.~\ref{fig8});
then, from these profiles, we derived the median values of
$R\rs{fs}/R\rs{cd}$ and compared them with those derived from runs
EX-C1.5-D1 and EX-C5.0-D2 (see Fig.~\ref{fig_append}). In both cases
analyzed, with ejecta clumps with either small size and low density
perturbations or large size and high density perturbations (namely the
two limit cases explored in this paper), we found that the median values
of $R\rs{fs}/R\rs{cd}$ derived from models with different initial ages are
consistent within the error bars, the value being slightly lower (higher)
in the model with initial age of the remnant $t\rs{snr0}=2$~yr than in
the model with $t\rs{snr0}=20$~yr when the clump size is 1\% (2\%) and
the maximum density perturbation is $\nu = 1.5$ ($\nu = 5$). We conclude
therefore that the results presented here do not depend on the initial
age of the simulated remnant. Indeed our results undoubtedly show
that the average separation between the contact discontinuity and the
forward shock strongly depends on the clumpy structure of the ejecta and,
in particular, on the size and density contrasts of the clumps.

\end{document}